\documentclass{article}
\usepackage[T1]{fontenc}
\usepackage{microtype}
\usepackage{graphicx}
\usepackage{booktabs}
\usepackage{tabularx}
\usepackage{array}

\usepackage[accepted]{icml2025}
\usepackage{amsmath,amssymb,mathtools}
\usepackage{placeins}
\usepackage{xcolor}
\usepackage[hyperfootnotes=false]{hyperref}
\setcitestyle{numbers,square,sort&compress}
\makeatletter
\renewcommand{\printAffiliationsAndNotice}[1]{%
  \begingroup
  \renewcommand{\thefootnote}{}%
  \footnotetext{\raggedright Correspondence to: \icmlcorrespondingauthor@text.}%
  \endgroup
}
\makeatother
\hypersetup{colorlinks=true,citecolor=blue,linkcolor=blue,urlcolor=blue}
\newcolumntype{Y}{>{\raggedright\arraybackslash}X}
\newcommand{\code}[1]{\texttt{#1}}
\icmltitlerunning{RoutingTrie: Metadata Routing for Constrained Generative Retrieval}

\begin{document}
\twocolumn[
\icmltitle{RoutingTrie: Metadata Routing for Constrained Generative Retrieval}
\begin{icmlauthorlist}
\icmlauthor{Darshan Fofadiya}{}
\icmlauthor{Yang Liu}{}
\icmlauthor{Shuying Sun}{}
\icmlauthor{Kyumin Park}{}
\icmlauthor{Juhana Kangaspunta}{}
\par\vspace{0.5em}
\href{https://www.shopify.com}{\includegraphics[width=24mm]{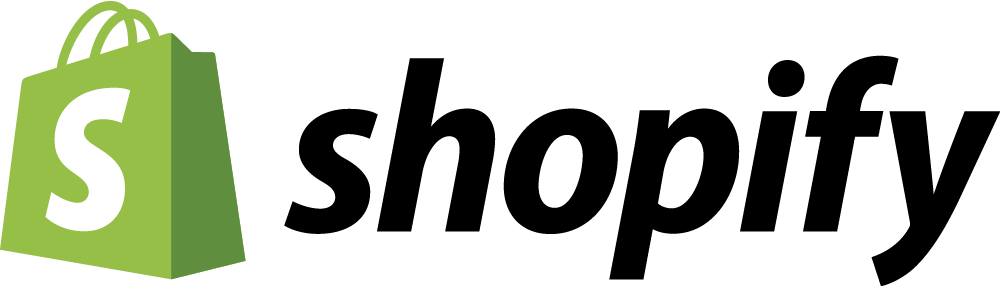}}
\end{icmlauthorlist}
\icmlcorrespondingauthor{Shuying Sun}{shuying.sun@shopify.com}
\icmlkeywords{generative retrieval, metadata constraints, constrained decoding, tries, recommender systems}
\vskip 0.2in
]
\printAffiliationsAndNotice{}

\begin{abstract}

Advances in large language models and semantic IDs (SIDs) have brought generative retrieval (GR) into industrial applications. These systems must return items that satisfy request metadata constraints as well as relevance requirements. A catalog trie ensures SID validity, but filtering metadata only after generation wastes computation on rejected candidates and may leave too few eligible results. We propose RoutingTrie to incorporate metadata constraints before beam selection. RoutingTrie maintains one shared SID trie and builds a compact index for each metadata field using Dense Bitset or Sparse List. We derive a storage cost formula to guide the choice between these index representations for different field distributions. During decoding, field probes determine whether requested values occur below candidate edges and produce a combined mask that guides beam selection. Experiments on ShopTrie, our configurable synthetic benchmark, demonstrate compact storage and efficient CPU and GPU probing across diverse metadata distributions. In a production replay, RoutingTrie raises Hit@100 from 67.5\% to 96.1\%, a gain of 28.6 percentage points, with only 1.1 ms latency increase. We will release the ShopTrie generator, benchmark scripts, and RoutingTrie CPU and Triton GPU implementations to support reproducibility.
\end{abstract}

\begin{figure}[t]
\centering
\includegraphics[width=\linewidth, trim={6 4 5 5}, clip]{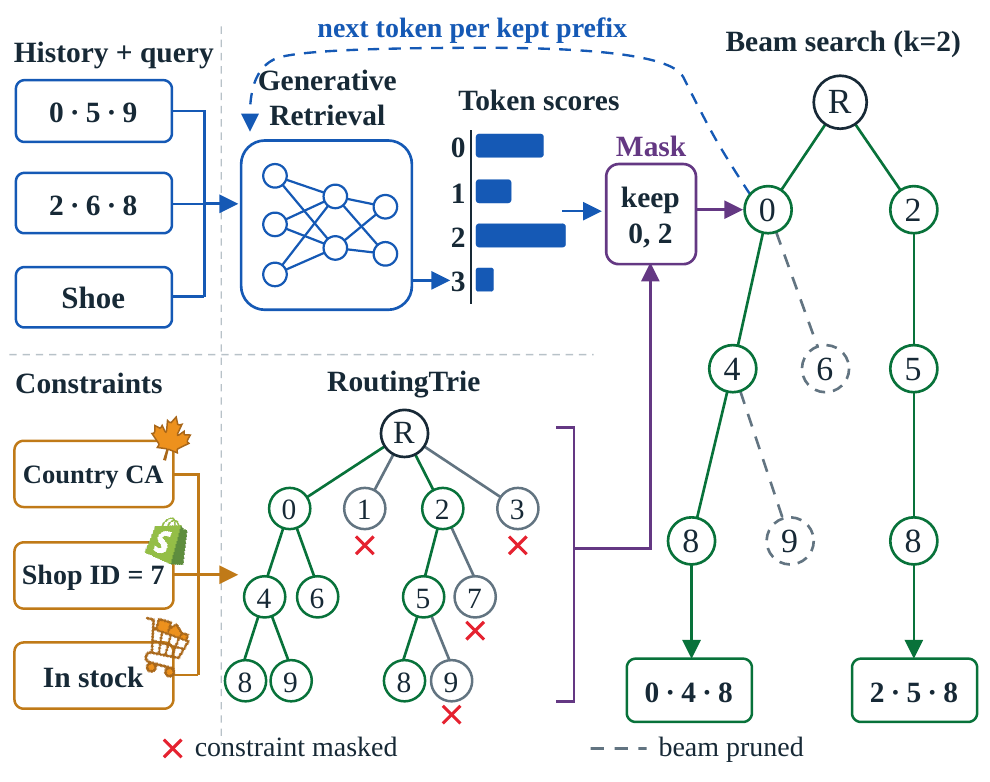}
\vspace{-20px}
\caption{\textbf{Constraint aware generative retrieval.} The GR model produces SID token scores from user history SIDs and an optional query. RoutingTrie uses metadata indexes to check requested values below candidate edges and combines these checks into a mask. Beam search retains the two highest scoring remaining prefixes, and resolved items are checked against the full request.}
\vspace{-10px}
\label{fig:groverview}
\end{figure}

\section{Introduction}
\label{sec:introduction}

Generative retrieval (GR) has gained momentum with advances in large language models and semantic IDs (SIDs). By encoding items as discrete token sequences, SIDs enable models to generate item identifiers directly from user context \citep{rajput2023tiger,si2023seater,zheng2024lcrec}. This approach has progressed from research models to industrial search and recommendation systems operating over large catalogs \citep{zhou2025onerec,he2025plum}. Such applications require GR to retrieve relevant items while respecting the metadata constraints of individual search and recommendation requests.

GR retrieves items by generating SIDs, which must correspond to entries in the catalog. Since not every token sequence forms a valid SID, a catalog trie of valid SIDs constrains the available continuations during decoding \citep{decao2021genre,su2026static,anandakumar2026flashtrie}. Figure~\ref{fig:groverview} illustrates this trie-based serving flow. The model receives the SIDs of items in a user's interaction history and an optional search query, then produces scores for candidate SID tokens. At each SID level, the trie limits valid continuations, and beam search uses these scores to retain the highest scoring prefixes. Completed SIDs are then resolved to items.

A valid SID, however, does not ensure that the item satisfies the request. In search and recommendation, a shopper may want only products available for purchase in Canada or items from a particular shop. Such requests impose metadata constraints on the serving flow shown in Figure~\ref{fig:groverview}. A straightforward approach generates SIDs first and post filters the resolved items afterward. This spends model computation on candidates that are later discarded. These candidates also consume limited beam capacity, potentially excluding eligible items before filtering \citep{wu2025sigir}. As a result, the system may return fewer items than requested even when enough eligible items exist in the catalog. This motivates applying metadata constraints before beam selection.

We introduce RoutingTrie to apply metadata constraints during SID generation in GR. As shown in Figure~\ref{fig:groverview}, it builds a separate memory efficient index for each metadata field over one shared SID trie. During decoding, these indexes identify whether requested values occur below candidate edges. Their results are combined into a mask before beam selection, helping preserve beam capacity for eligible items.
We evaluate RoutingTrie with our synthetic benchmark ShopTrie and our production GR system. ShopTrie experiments show compact storage and efficient CPU and GPU probing across diverse metadata distributions. In a production replay, RoutingTrie raises Hit@100 from 67.5\% to 96.1\%, a gain of 28.6 percentage points with a latency increase of only 1.1 ms.

We make three contributions.
\begin{enumerate}
\item We introduce RoutingTrie for production friendly low latency metadata filtering during SID generation in GR, supporting fields with low and high cardinality on one shared trie.
\item We provide a formula for comparing storage costs and selecting a suitable index representation for each metadata field.
\item We support reproducible comparisons through ShopTrie, a configurable synthetic benchmark, and will release its generator, benchmark scripts, and RoutingTrie CPU and Triton GPU kernels.
\end{enumerate}

\section{Related work}
\label{sec:related}

\paragraph{Generative retrieval.} Unlike dense retrieval, which matches query or user representations against candidate representations \citep{karpukhin2020dpr,xiong2021ance,yi2019sampling}, GR selects candidates by generating item or document identifiers.
TIGER and SEATER generated structured item identifiers from user histories \citep{geng2022p5,rajput2023tiger,si2023seater}. LC-Rec incorporates collaborative semantics into language model based recommendation, and LIGER studies a hybrid of generative and dense retrieval \citep{zheng2024lcrec,yang2024liger,tu2026revisiting}. Sequential models such as GRU4Rec, SASRec, BERT4Rec, and HSTU provide ways to represent user histories with dense item ids \citep{hidasi2016gru4rec,kang2018sasrec,sun2019bert4rec,zhai2024hstu,liu2026t}. OneRec and its subsequent reports develop large scale generative recommendation systems, while PLUM describes industrial SID retrieval \citep{deng2025onerec,zhou2025onerec,zhou2025onerecv2,he2025plum}. OneRec-Think and OneReason study reasoning within generative recommendation \citep{liu2025onerecthink,onerecteam2026onereason}. Beyond GR, SIDs also appear in ranking, while GR2 uses them in the later reranking stage \citep{singh2024semanticids,li2026gr2}. %

\paragraph{Valid generation with tries.} In GR, not all SID sequences are valid, so a trie restricts decoding to valid prefixes.
Work on valid path search examines how corpus constraints interact with beam search \citep{wu2025sigir}. OneGR explores this search using weighted automata and one pass scoring \citep{wang2026onegr}. For efficient execution, DISC accelerates prefix checks, while STATIC and FlashTrie optimize valid path traversal on accelerators \citep{ye2025disc,su2026static,anandakumar2026flashtrie}.
These methods improve generation and traversal, while this paper tries to solve metadata constraints for an existing trie.

\paragraph{Metadata filtering.} Search and recommendation systems in production must combine relevance with predicates over attributes such as labels, regions, and prices. 
Filtered vector retrieval supports predicates through graph traversal \citep{gollapudi2023filtered,wu2022hqann,patel2024acorn,cai2024ung}, partitions or shared and specialized indexes \citep{gupta2023caps,jin2026curator,li2025sieve}, and numeric range handling \citep{engels2024window,liang2024unify}. Execution research covers relational integration and GPU processing \citep{zhang2023vbase,xi2025vecflow}, while benchmarking examines selectivity and execution strategy \citep{shi2025fannsbenchmark}. For value sets, Bloom filters provide approximate membership checks, while Roaring bitmaps compress exact sets \citep{bloom1970,lemire2016roaring}.
These techniques address metadata selectivity and representation in dense retrieval, while this paper addresses a gap in metadata filtering for GR.

\section{Problem definition}
\label{sec:problem}

\begin{figure}[t]
\centering
\includegraphics[width=\linewidth, trim={5 0 2 14}, clip]{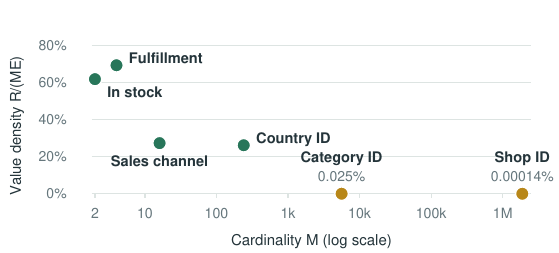}
\vspace{-20px}
\caption{\textbf{Two metadata patterns based on real distributions.} Low cardinality fields show high value density (\textcolor{green!50!black}{green dots}), while high cardinality fields show low value density (\textcolor{yellow!70!black}{yellow dots}). }%
\vspace{-10pt}
\label{fig:constraint-reuse}
\end{figure}

Our goal is to support metadata filtering during SID generation in GR. The input consists of a catalog trie of valid SIDs with $E$ edges, item metadata, and a request $C$. A request combines atomic conditions $(f,c)$ using AND, OR, and NOT, where $f$ is a metadata field and $c$ is a value. For example, $C$ may require country ID 25 and shop ID 7. An item is eligible if its metadata satisfies the Boolean expression specified by $C$.

The output is a mask over the candidate next SID tokens of each live beam prefix. For a candidate edge $e$, $\operatorname{Mask}(e,C)$ is 1 if the edge remains available for beam selection and 0 otherwise. As Figure~\ref{fig:groverview} shows, the mask acts on model scores during beam selection. Without metadata constraints, all trie valid candidate edges remain available.

For each metadata field, $M$ denotes its cardinality, with values encoded as integers from $0$ to $M-1$. Each edge corresponds to a set of distinct field values found below it, and $R$ is the sum of these set sizes across all $E$ edges. Value intensity $I=R/E$ measures the average number of values below an edge, while value density $R/(ME)$ measures the average fraction of possible values present below an edge.

In this problem, real catalog metadata varies sharply in cardinality and value density, with low cardinality fields showing high value density and high cardinality fields showing low value density. Figure~\ref{fig:constraint-reuse} illustrates these patterns across six representative fields with cardinalities from 2 to 2M in real metadata distributions. 
Value density
exceeds 50\% for binary fields and falls below 0.1\% for category and shop IDs.

\section{Methodology}
\label{sec:method}

We present RoutingTrie to support metadata filtering during SID generation in GR. It indexes each metadata field separately and combines the requested fields' probe results into a mask before beam selection, as shown in Figure~\ref{fig:method}.

\subsection{RoutingTrie overview}
\label{sec:overview}

\begin{figure}[t]
\centering
\includegraphics[width=\linewidth, trim={2 2 2 2}, clip]{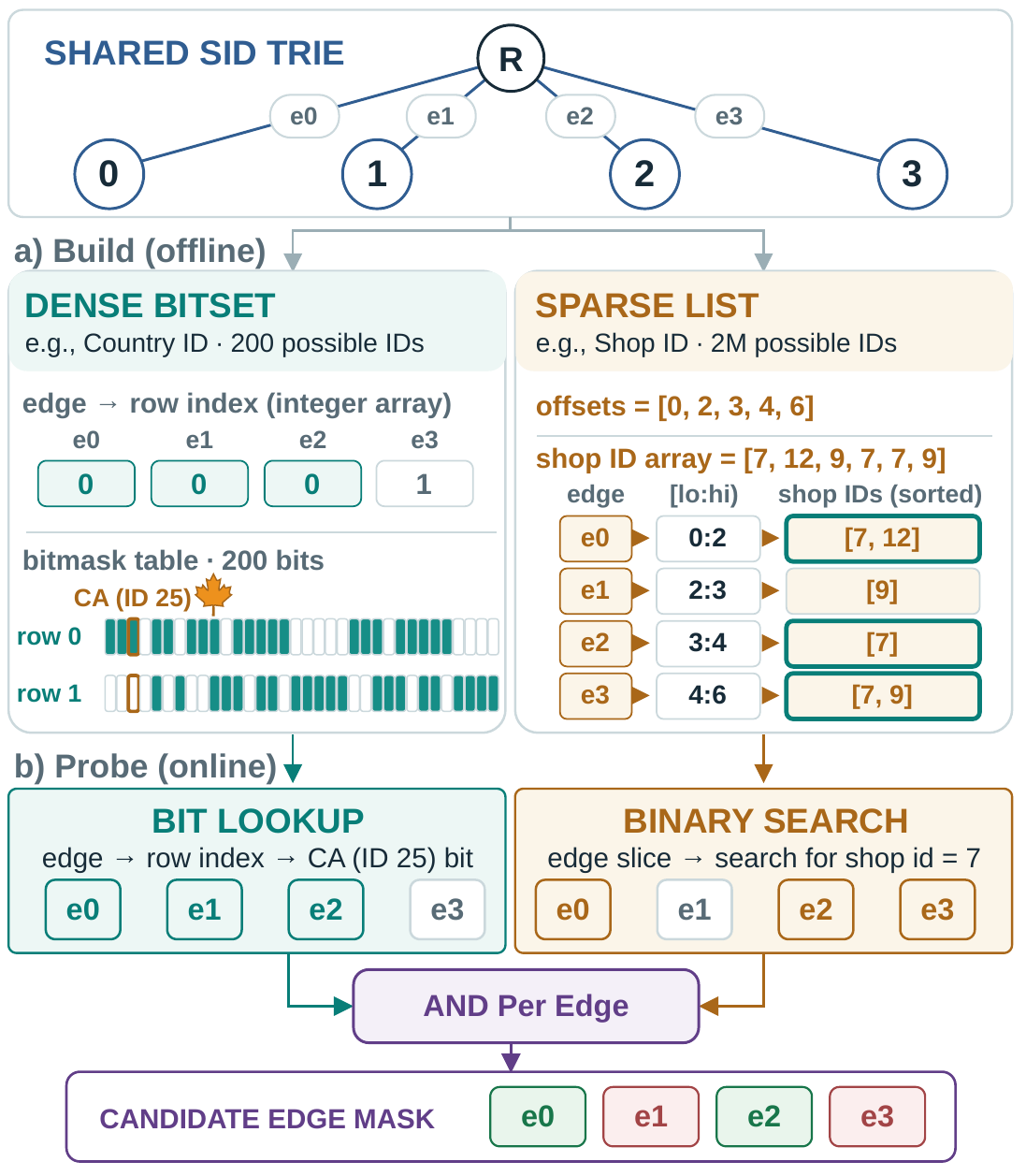}
\vspace{-19px}
\caption{\textbf{RoutingTrie overview.} RoutingTrie first builds metadata field indexes on one shared SID trie (a). During decoding, it checks requested metadata values by bit lookup or binary search, then combines the results to form the candidate edge mask (b).}
\label{fig:method}
\vspace{-10px}
\end{figure}

RoutingTrie adds metadata filtering to a shared SID trie independently of the trie traversal algorithm.  As shown in Figure~\ref{fig:method}, it first builds a separate index for each field, recording the distinct values among items below each edge. During decoding, trie traversal enumerates child edges of the current beam prefixes. Field probes then check the requested values in the corresponding metadata indexes and combine the results into a mask before beam selection.

The metadata index construction begins by aggregating each field's values from the leaves upward. Leaf entries include values from all items sharing an SID, and parent entries are the unions of their child sets. For each field, \code{edgeValues[e]} records the distinct values below edge $e$, such as the shop values $\{7,12\}$ for $e_0$ in Figure~\ref{fig:method}(a). We then store these sets in a separate field index. To reduce memory use across the patterns in Figure~\ref{fig:constraint-reuse}, we use Dense Bitset for low cardinality fields with high value density and Sparse List for high cardinality fields with low value density. 

At each decoding step, RoutingTrie checks the requested metadata values on the child edges of every live beam prefix. Each field uses its own index to return $\operatorname{Probe}_f(e,c)$, which is $1$ when requested value $c$ appears below edge $e$ and $0$ otherwise. %
The probes can run in parallel, with each producing a Boolean mask over the candidate edges. 
RoutingTrie evaluates the Boolean expression specified by the request over these masks, using AND, OR, and NOT to determine which edges remain available for beam selection.
\[
\operatorname{Mask}(e,C)
=
C\!\left[(f,c)\mapsto\operatorname{Probe}_f(e,c)\right].
\]
For example, Figure~\ref{fig:method}(b) shows masks $(1,1,1,0)$ for country ID 25 and $(1,0,1,1)$ for shop ID 7 over $e_0$ through $e_3$. Their intersection, $(1,0,1,0)$, keeps $e_0$ and $e_2$ available for beam selection. %
Since an SID may represent multiple items with different metadata, passing the combined mask does not replace the final item eligibility check after SID resolution.

\subsection{Dense Bitset metadata index and probe}
\label{sec:dense}

Dense Bitset handles compact fields when many edges share the same set of reachable values. For a field such as country, the possible values fit in a short bitset with one bit per country ID. Testing a requested ID then requires only a bit lookup. Storing a separate bitset for every edge would repeat the same pattern when edges reach identical sets of countries. Dense Bitset therefore stores each distinct row once and uses an integer array to map edges to rows. In Figure~\ref{fig:method}(a), $e_0,e_1,e_2$ share country row 0, while $e_3$ uses row 1. This mapping preserves direct access by edge ID without copying repeated rows.

Dense construction converts each edge set into a bitset and stores identical rows only once. Algorithm~\ref{alg:dense} specifies this build stage and the matching probe. \textbf{BuildDense} takes the $E$ sets in \code{edgeValues} for one field and its cardinality $M$. It returns a table \code{rows} of distinct bitsets and an array \code{rowOf} that identifies the row for each edge. A row has $W=\lceil M/64\rceil$ words of 64 bits each, enough to assign one bit to every possible value. For each edge $e$, the algorithm starts with a zero row and sets the bits for the values in \code{edgeValues[e]}. If an identical row already exists, \code{rowOf[e]} points to it. Otherwise, the new row is added to the table. Rows must be compared in full before sharing them, even if a fingerprint narrows the search, to avoid incorrect probe results. 
With $D$ distinct rows, Dense Bitset uses $4E+8DW$ bytes to store the edge mapping and bitset rows. Row share is $D/E$, with lower values indicating more row sharing.

\begin{algorithm}[b]
\caption{Dense Bitset construction and probe}
\label{alg:dense}
\small
\begin{algorithmic}[1]
\STATE \textbf{BuildDense}($\mathit{edgeValues},M,E$)
\STATE $W\gets\lceil M/64\rceil$, $\mathit{rows}\gets[\,]$
\FOR{$e=0$ to $E-1$}
  \STATE $b\gets\operatorname{zeros}(W)$
  \FOR{$\mathit{value}\in\mathit{edgeValues}[e]$}
    \STATE $\operatorname{SetBit}(b,\mathit{value})$
  \ENDFOR
  \STATE $\mathit{rowOf}[e]\gets\operatorname{InternExact}(\mathit{rows},b)$
\ENDFOR
\STATE \textbf{return} $\operatorname{DenseIndex}(\mathit{rows},\mathit{rowOf},M)$
\STATE \textbf{ProbeDense}($\mathit{dense},e,\mathit{value}$)
\IF{$\mathit{value}<0$ or $\mathit{value}\ge\mathit{dense}.M$}
  \STATE \textbf{return} false
\ENDIF
\STATE $r\gets\mathit{dense}.\mathit{rowOf}[e]$
\STATE $w\gets\mathit{dense}.\mathit{rows}[r,\lfloor\mathit{value}/64\rfloor]$
\STATE \textbf{return} $((w\gg(\mathit{value}\bmod64))\mathbin{\&}1)==1$
\end{algorithmic}
\vspace{-2px}
\end{algorithm}

Dense probing follows the edge to its row and tests one bit. \textbf{ProbeDense} takes the built metadata index, an edge $e$, and a requested value $c$, and returns whether $c$ appears below that edge.
For a valid field value $c$, the probe uses \code{rowOf[e]} to locate the bitset row and tests bit $c\bmod64$ in word $\lfloor c/64\rfloor$.
In Figure~\ref{fig:method}(b), country ID 25 is present in row 0 and absent from row 1, so $e_0,e_1,e_2$ pass and $e_3$ fails. The probe takes \(O(1)\) time per edge through two dependent array reads, regardless of $M$ or the number of countries below the edge.

\subsection{Sparse List metadata index and probe}
\label{sec:sparse}

Sparse List handles high cardinality fields whose values reach relatively few edges. Shop ID is one such field in Figure~\ref{fig:constraint-reuse}. Its cardinality is roughly two million, so every distinct bitset row would reserve roughly two million bits even for an edge with one shop. Most bit positions would be unused on average. Sparse List instead stores only values that occur below each edge. It concatenates those values in one array and records where each edge's segment begins and ends in an array of $E+1$ offsets. As in compressed sparse row storage, this layout needs no separate value array or pointer for each edge.

Sparse construction stores each edge's sorted values in one shared array, so memory follows the number of values stored across edges. Algorithm~\ref{alg:sparse} specifies this build stage and the matching probe. \textbf{BuildSparse} takes the $E$ sets in \code{edgeValues} for one field and its cardinality $M$. It returns a value array \code{values} with sorted segments for each edge and an offset array \code{offsets}. For each edge, it sorts the edge's set, appends it to \code{values}, and records the new array length as the next offset. 
The values for edge $e$ start at \code{offsets[e]} and stop before \code{offsets[e+1]}. 
In Figure~\ref{fig:method}(a), the offsets $[0,2,3,4,6]$ divide the shop ID array $[7,12,9,7,7,9]$ into four edge segments. Edge $e_0$ contains $[7,12]$ and $e_1$ contains $[9]$. 
An edge stores empty set when its start and end offsets are equal. 
Sorting at build time permits binary search during decoding. 
If the completed value array has $R$ entries, it stores $I$ values per edge on average. With 32 bit values and 64 bit offsets, the sparse list metadata index occupies \(4R+8(E+1)\) bytes. For IDs that fit in 32 bits, this count depends on $R$ and $E$ rather than directly on $M$.

\begin{algorithm}[!tb]
\caption{Sparse List construction and probe}
\label{alg:sparse}
\small
\begin{algorithmic}[1]
\STATE \textbf{BuildSparse}($\mathit{edgeValues},M,E$)
\STATE $\mathit{values}\gets[\,]$, $\mathit{offsets}[0]\gets0$
\FOR{$e=0$ to $E-1$}
  \STATE $\operatorname{Extend}(\mathit{values},\operatorname{Sort}(\mathit{edgeValues}[e]))$
  \STATE $\mathit{offsets}[e+1]\gets|\mathit{values}|$
\ENDFOR
\STATE \textbf{return} $\operatorname{SparseIndex}(\mathit{values},\mathit{offsets},M)$
\STATE \textbf{ProbeSparse}($\mathit{sparse},e,\mathit{value}$)
\IF{$\mathit{value}<0$ or $\mathit{value}\ge\mathit{sparse}.M$}
  \STATE \textbf{return} false
\ENDIF
\STATE $\mathit{lo},\mathit{hi}\gets\mathit{sparse}.\mathit{offsets}[e],\mathit{sparse}.\mathit{offsets}[e+1]$
\STATE $j\gets\operatorname{lower\_bound}(\mathit{sparse}.\mathit{values},\mathit{lo},\mathit{hi},\mathit{value})$
\IF{$j=\mathit{hi}$}
  \STATE \textbf{return} false
\ENDIF
\STATE \textbf{return} $\mathit{sparse}.\mathit{values}[j]==\mathit{value}$
\end{algorithmic}
\vspace{-2px}
\end{algorithm}

Sparse probing searches only the sorted segment of the candidate edge. \textbf{ProbeSparse} takes the built index, edge $e$, and requested value $c$, and returns whether $c$ occurs in that segment.
For a valid field value $c$, the probe uses the two offsets to locate the edge's segment. It then uses \code{lower\_bound} to find the first position whose value is at least $c$, returning the end offset if no such value exists.
The probe returns true only if this position is still inside the segment and holds $c$. 
In Figure~\ref{fig:method}(b), shop ID 7 passes $e_0,e_2,e_3$ and fails $e_1$. For a segment of $s$ values, the search takes \(O(\log(s+1))\) comparisons. Its cost depends on the number of values below the edge rather than the field cardinality $M$.

\subsection{Choosing a representation}
\label{sec:choice}

\begin{figure}[!tb]
\centering
\includegraphics[width=\linewidth, trim={4 8 2 4}, clip]{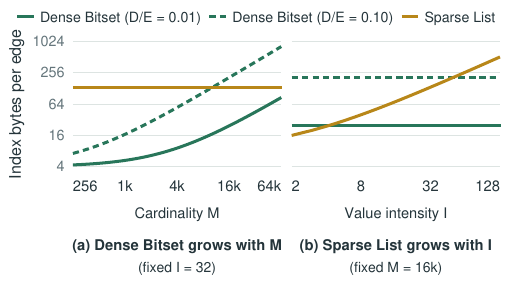}
\vspace{-20px}
\caption{\textbf{Metadata patterns guide index choice.} Increasing cardinality raises Dense Bitset memory while Sparse List stays constant (a). Increasing value intensity raises Sparse List memory while Dense Bitset stays constant (b). Each comparison holds the other factors fixed.}
\vspace{-8px}
\label{fig:storage-scaling}
\end{figure}

RoutingTrie chooses an index for each field to reduce memory use. Figure~\ref{fig:storage-scaling} shows the two storage costs at fixed row share. Dense Bitset grows with the cardinality $M$ in panel (a), while Sparse List grows with value intensity $I$ in panel (b). Sparse List saves space when cardinality is high but each edge contains few values (low value intensity). Dense Bitset becomes more attractive when cardinality is low or the edge lists are long (high value intensity). %

To make the index choice explicit, we derive a selection rule from the memory footprints of Dense Bitset and Sparse List. Since both use the same trie, we compare only their field arrays. Dense Bitset uses $4E+8D\lceil M/64\rceil$ bytes for $D$ distinct rows, while Sparse List uses $4R+8(E+1)$ bytes for $R$ stored values across its edge segments. Comparing these costs, Dense Bitset uses less memory when
\begin{equation}
I>2\frac{D}{E}\left\lceil\frac{M}{64}\right\rceil-1-\frac{2}{E}.
\label{eq:main}
\end{equation}
Sparse List uses less memory below this threshold, and both indexes use the same memory at equality. This threshold assumes 32 bit row indices and value IDs, with 64 bit bitset words and offsets. For optimized layouts, we apply the same memory comparison using the widths and counts of the arrays actually stored.

\begin{figure}[!tb]
\centering
\includegraphics[width=\columnwidth, trim={2 4 2 2}, clip]{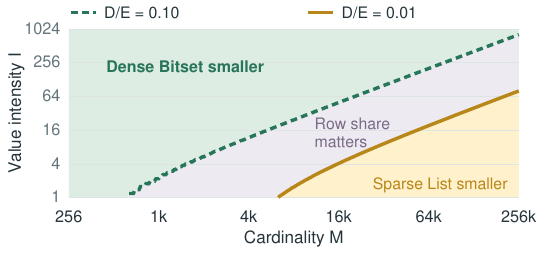}
\vspace{-20px}
\caption{\textbf{Row sharing lowers the memory threshold.} The curves show where the two indexes use equal memory at different row share values.}
\vspace{-10px}
\label{fig:storage-crossover}
\end{figure}

More row sharing lowers $D/E$ and the threshold, allowing Dense Bitset to use less memory than Sparse List at smaller $I$. Figure~\ref{fig:storage-crossover} illustrates this effect as cardinality $M$ varies. For example, at $M=16{,}384$ and $E=100{,}000$, reducing $D/E$ from 0.10 to 0.01 lowers the threshold on $I$ from about 50 to 4. When memory costs are similar, probe latency for the target beam width and device can guide the final choice.

\section{Experiment}
\label{sec:evaluation}

Our evaluation has two parts. We use our synthetic benchmark ShopTrie for reproducible measurements of index construction cost, memory use, and probe latency. We then assess RoutingTrie using data from our inhouse production system, measuring end-to-end GR hit rate and request latency in deployment.

\subsection{ShopTrie benchmark}
\label{sec:shoptrie-benchmark}

\paragraph{Setup.}\label{sec:shoptrie-setup}

We introduce ShopTrie to compare metadata indexes and probing methods on common, reproducible inputs. Real GR systems differ in SID assignments, trie structures, and metadata distributions, so ShopTrie generates a deterministic trie and metadata from a configuration and seed. This allows different methods to be evaluated on identical inputs. 
We configure ShopTrie with three SID levels, 8,192 possible tokens per level, and eight metadata fields with cardinalities from two to roughly two million. Selected metadata distributions are calibrated at the leaf level using real catalog summaries. For each field, leaf edges store the union of their items' values, and parent edges store the union of their child sets. We use the same configuration with seeds 42 through 46 to generate different tries and metadata assignments. The resulting data statistics appear in the Count and Values/edge columns of Table~\ref{tab:shoptrie-main}.

ShopTrie primarily measures metadata index size and probe latency on CPU and GPU. To isolate probe latency, we use an untimed beam trace to record candidate edges at each SID level. We then time only the metadata checks and Boolean mask generation on these fixed edges, excluding child expansion, scoring, and beam selection. We report mean and p95 latency from at least 10,000 probe calls in every seed to capture average and tail performance and inform deployment decisions. We also record index construction time for reference. All the experiments are run on an Intel Xeon CPU with 16 cores and an NVIDIA B300 GPU. The implementation uses NumPy and PyTorch, with GPU kernels written in Triton.

\begin{table*}[!t]
\centering
\vspace{-6px}
\caption{ShopTrie index build, GPU resident size, and metadata probe latency at beam width 64 over five seeds. Values per edge are shown as mean/p50, with the mean equal to value intensity $I$. Latencies are mean/p95 in milliseconds per SID layer and across all three levels.}
\label{tab:shoptrie-main}
\footnotesize
\setlength{\tabcolsep}{2.0pt}
\renewcommand{\arraystretch}{1.12}
\begin{tabular*}{\textwidth}{@{\extracolsep{\fill}}llrlrrcccc@{}}
\toprule
& & & & & & \multicolumn{2}{c}{CPU (ms)} & \multicolumn{2}{c}{B300 GPU (ms)} \\
\cmidrule(lr){7-8}\cmidrule(l){9-10}
Component & Count & Values per edge & Encoding & Build (s) & Memory (GiB) & Layer & Total & Layer & Total \\
\midrule
Shared trie & 338M edges & -- & CSR topology & 6.8 & 0.77 & -- & -- & -- & -- \\
\addlinespace[2pt]
In stock & 2 & 1.24/1 & Dense bitset & 7.8 & 0.31 & 0.03/0.07 & 0.10/0.11 & 0.03/0.03 & 0.08/0.09 \\
On sale & 2 & 1.28/1 & Dense bitset & 7.2 & 0.31 & 0.03/0.07 & 0.10/0.11 & 0.03/0.03 & 0.08/0.09 \\
Condition & 3 & 1.35/1 & Dense bitset & 11.7 & 0.31 & 0.03/0.07 & 0.10/0.11 & 0.03/0.03 & 0.08/0.09 \\
Sales channel & 16 & 4.38/4 & Dense bitset & 8.9 & 0.63 & 0.04/0.09 & 0.12/0.14 & 0.03/0.03 & 0.08/0.09 \\
Fulfillment & 4 & 2.78/3 & Dense bitset & 6.5 & 0.31 & 0.03/0.07 & 0.10/0.11 & 0.03/0.03 & 0.08/0.09 \\
Country ID & 240 & 62.95/62 & Dense bitset & 34.7 & 1.33 & 0.09/0.19 & 0.26/0.28 & 0.03/0.03 & 0.08/0.09 \\
\addlinespace[2pt]
Category ID & $\approx$5.6K & 1.38/1 & Sparse list & 224.0 & 1.01 & 0.22/0.55 & 0.67/0.70 & 0.03/0.04 & 0.08/0.10 \\
Shop ID & $\approx$1.9M & 2.71/1 & Sparse list & 90.9 & 5.92 & 0.42/0.92 & 1.26/1.31 & 0.03/0.04 & 0.09/0.10 \\
\midrule
Total resident memory & -- & -- & -- & -- & 10.91 & -- & -- & -- & -- \\
\bottomrule
\end{tabular*}
\par
\vspace{-10px}
\end{table*}

\paragraph{Results.}\label{sec:shoptrie-main}
Table~\ref{tab:shoptrie-main} summarizes the ShopTrie results at beam width 64. We highlight four main findings.

\emph{The shared trie and selected metadata indexes occupy 10.91 GiB of memory.} This footprint reflects storage optimizations beyond the uniform layouts assumed in Equation~\ref{eq:main}. Our implementation reduces these costs by adapting storage widths to field cardinality and removing unnecessary mappings and offsets. For the five small Dense Bitset fields, each bitset fits in one or two bytes and is stored directly per edge, avoiding the row mapping counted by the $4E$ term. Category ID uses 16 bit values instead of the 32 bit values assumed by the $4R$ term. It also omits leaf offsets because each leaf edge contains exactly one category value in ShopTrie. These choices explain the smaller footprints reported in Table~\ref{tab:shoptrie-main}. Further savings are possible by packing the five small fields together. Packing their 27 bits into one 32 bit row per edge could reduce their combined size from 1.89 to 1.26 GiB and the total resident size to 10.28 GiB while retaining bit tests.

\emph{GPU probe latency remains similar across fields with very different cardinalities.} Mean total GPU probe latency ranges from 0.075 to 0.087 ms, although field cardinalities range from two to roughly two million. Dense Bitset tests one bit, while Sparse List searches only the values stored on each candidate edge. 
Short edge lists keep Sparse List probing efficient despite the additional steps of binary search compared with checking a single bit. The value intensity $I$ is only 1.38 for Category ID and 2.71 for Shop ID.
Selective masks also reduce the number of prefixes expanded at later SID levels.

\emph{GPU probing is faster and more stable than CPU probing.} It distributes independent edge probes across GPU threads and combines metadata checks and Boolean mask generation in one Triton kernel per field at each SID level. This exposes parallel work and reduces kernel launch overhead. Across all eight fields, GPU probing is about 1.31 to 14.48 times as fast as CPU probing, with mean latency ranging from 0.075 to 0.087 ms. 
Across all eight fields, GPU p95 stays within 0.018 ms of the mean, while the CPU gap reaches 0.046 ms.

\emph{Compact fields have short build times, which matters when indexes must be rebuilt.} 
Initial construction takes 6.5 to 11.7 s for the five compact Dense Bitset indexes and 34.7 s for Country ID. Category ID and Shop ID take longer, at 224.0 and 90.9 s. Once the indexes are built, changing stock availability does not require changing the SID trie. Our update implementation recomputes affected edge summaries and refreshes the corresponding field indexes before publishing them to the GPU. Updating 0.01\% of leaf sets takes 0.077 to 0.093 s for the compact fields, compared with 6.288 s for Category ID and 19.530 s for Shop ID. Compact bitsets can be patched in place, whereas the current Sparse List implementation repacks contiguous arrays when segment lengths change. More details and comparisons can be found in Table~\ref{tab:shoptrie-metadata-updates}.

\begin{table}[!htbp]
\centering
\caption{Metadata index update time on ShopTrie. Each field independently updates the indicated fraction of leaf metadata sets. Times in seconds include exact aggregation of affected ancestors, index updates and publication to the GPU.}
\label{tab:shoptrie-metadata-updates}
\footnotesize
\setlength{\tabcolsep}{2pt}
\begin{tabularx}{\columnwidth}{@{}Yrrrr@{}}
\toprule
Field & 0.01\% & 0.1\% & 1\% & 5\% \\
\midrule
In stock & 0.093 & 0.222 & 1.415 & 6.293 \\
On sale & 0.077 & 0.224 & 1.236 & 5.874 \\
Condition & 0.089 & 0.282 & 1.614 & 7.051 \\
Sales channel & 0.090 & 0.293 & 1.573 & 7.427 \\
Fulfillment & 0.085 & 0.280 & 1.472 & 7.240 \\
Country ID & 1.053 & 1.942 & 6.140 & 19.345 \\
Category ID & 6.288 & 7.664 & 14.499 & 23.608 \\
Shop ID & 19.530 & 22.496 & 31.335 & 53.427 \\
\midrule
Sum over eight fields & 27.305 & 33.404 & 59.283 & 130.265 \\
Replaced leaf edges & 31,884 & 318,843 & 3,188,435 & 15,942,174 \\
Affected middle edges & 31,700 & 301,730 & 2,253,913 & 6,769,635 \\
Affected root edges & 7,384 & 8,190 & 8,191 & 8,191 \\
\bottomrule
\end{tabularx}
\end{table}

\begin{figure}[!tb]
\centering
\includegraphics[width=\columnwidth, trim={1 2 2 2}, clip]{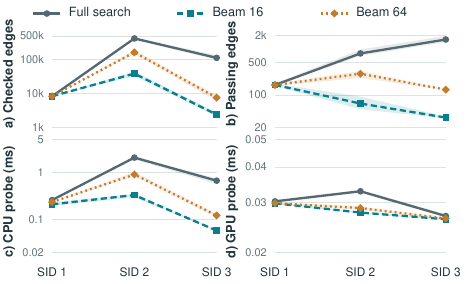}
\vspace{-18px}
\caption{\textbf{Shop ID edge flow across SID levels.} Panels (a) and (b) show checked and passing child edges. Panels (c) and (d) show per level CPU and GPU probe latency. Full search retains all passing prefixes. Lines show means and bands show ranges across seeds.}
\label{fig:shoptrie-edge-flow}
\vspace{-10px}
\end{figure}

\paragraph{SID level ablation.}\label{sec:shoptrie-sid-level}
The same beam width can produce different workloads across SID levels because retained prefixes have different numbers of children and metadata probes admit different numbers of edges. To understand how these differences affect latency, we evaluate a Shop ID query without a beam limit and with beam widths 16 and 64. At each level, we measure child edges checked, edges passing the Shop ID probe before beam selection, and probe latency on CPU and GPU, which are shown in Figure~\ref{fig:shoptrie-edge-flow}.

Figure~\ref{fig:shoptrie-edge-flow} shows that CPU probe latency responds strongly to the workload reduction from beam selection, while GPU latency varies much less. Beam 16 lowers CPU latency from 2.06 to 0.33 ms at SID 2 and from 0.665 to 0.057 ms at SID 3. The corresponding GPU reductions are much smaller, from 0.033 to 0.028 ms and from 0.027 to 0.026 ms. That is because GPU probes run in parallel, allowing large differences in edge count to produce small latency changes. With beam widths 16 and 64, GPU latency falls from about 0.030 ms at SID 1 to 0.028 and 0.029 ms at SID 2, respectively, despite checking more edges, then reaches about 0.026 ms at SID 3.

\begin{figure}[!tb]
\centering
\includegraphics[width=\columnwidth, trim={14 8 10 6}, clip]{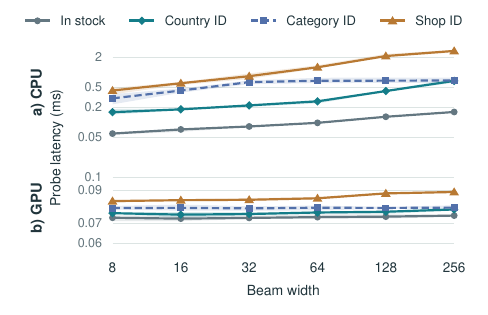}
\vspace{-20px}
\caption{\textbf{Probe latency across beam widths.} (a) CPU probe latency grows at different rates across fields. (b) GPU probe latency remains relatively flat across beam widths. Lines average medians across seeds, and bands show the seed range.}
\label{fig:shoptrie-beam-latency}
\vspace{-10px}
\end{figure}

\paragraph{Beam width ablation.}\label{sec:shoptrie-beam-width}
Beam width sets an upper limit on retained prefixes, while metadata filtering determines how much of that capacity is used. This ablation examines how the interaction affects probe latency on CPU and GPU. We vary beam width from 8 to 256 for four individual fields, where its CPU and GPU probe latency is shown in Figure~\ref{fig:shoptrie-beam-latency}.

Figure~\ref{fig:shoptrie-beam-latency} shows that increasing beam width has a much larger effect on CPU probe latency than on GPU probe latency. From width 8 to 256, CPU latency rises from 0.16 to 0.67 ms for Country ID and from 0.43 to 2.67 ms for Shop ID. The corresponding GPU increases are only about 3\% and 7\%, respectively. The CPU increase reflects the additional candidate work admitted by wider beams. In stock, Country ID, and Shop ID retain more prefixes as the beam widens, increasing the number of edges checked at later SID levels. Category ID instead levels off near 0.63 to 0.68 ms from width 32 onward because few prefixes pass at an early SID level, limiting further expansion. Beam width therefore has a substantial CPU latency cost when the request leaves enough prefixes for further expansion, while GPU latency changes much less over the tested range.

\begin{figure}[!tb]
\centering
\includegraphics[width=\columnwidth, trim={2 2 2 2}, clip]{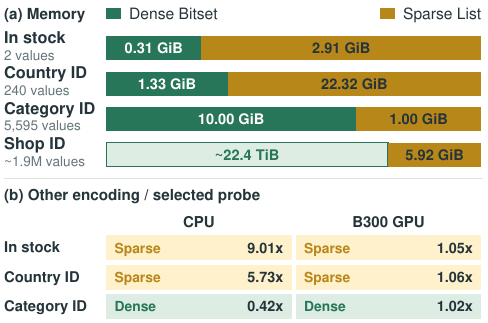}
\vspace{-20px}
\caption{\textbf{Memory and probe latency across index encodings.} (a) Resident index sizes, with logarithmic scaling within each field and an estimated Dense Bitset size for Shop ID. (b) Alternative encoding median probe latency divided by selected encoding latency on CPU and GPU. Ratios below one indicate a faster alternative.}
\vspace{-12px}
\label{fig:shoptrie-index-memory}
\end{figure}

\paragraph{Index choice ablation.}\label{sec:shoptrie-index-choice}
The main table uses different encodings for fields with different metadata, making their effects difficult to separate. To isolate the effect of index choice, we build both Dense Bitset and Sparse List for In stock, Country ID, and Category ID. Within each field, both encodings use the same query value, scores, and beam width 64. Figure~\ref{fig:shoptrie-index-memory} compares resident index size and total probe latency on CPU and GPU.

Matching the encoding to each field avoids substantial memory overhead. Figure~\ref{fig:shoptrie-index-memory}(a) shows that Sparse List pays heavily for offsets and repeated values, using about 9 times as much memory as Dense Bitset for In stock and 17 times for Country ID. Dense Bitset pays for possible values even when they are absent, making it about 10 times larger than Sparse List for Category ID. This cost becomes especially large for Shop ID, whose Dense Bitset size is estimated at 22.4 TiB, $\sim$3,900 times the measured Sparse List size.

Extra index memory does not guarantee lower probe latency, especially on GPU. As Figure~\ref{fig:shoptrie-index-memory}(b) shows, on CPU, Dense Bitset uses indexed bit access, while Sparse List can require repeated comparisons during binary search. For In stock and Country ID, Dense Bitset is about 9 and 5.7 times as fast as Sparse List, respectively, although Sparse List uses about 9 and 17 times as much memory. For Category ID, low value intensity $I$ keeps sparse lists short, narrowing Dense Bitset's CPU speedup to about 2.4 times at about 10 times the memory cost. On GPU, independent edge probes run in parallel using fused Triton kernels. The measured differences are much smaller, with Sparse List taking only about 5\% and 6\% longer for In stock and Country ID, respectively, and Dense Bitset taking about 2\% longer for Category ID.

\subsection{Production benchmark}
\label{sec:real-eval}

\paragraph{Setup.}\label{sec:real-setup}

ShopTrie measures index construction cost, memory use, and probe latency. To evaluate RoutingTrie in a real GR system, we replay historical purchase sessions using our production system. The system uses a 14B parameter model and a three level SID trie with 8,192 tokens per level. Evaluation uses a batch size of one and a beam width of 100. We replay 23,167 sessions, using each purchased item as the retrieval target and its shop ID as the metadata constraint. The shop ID field has multi-million cardinality.

Our production benchmark evaluates retrieval quality and deployment request latency separately. Retrieval quality is measured by item level Hit@100, defined as the fraction of sessions whose purchased item appears among the first 100 eligible items after SID resolution. Deployment request latency measures the complete GR pipeline running in TensorRT-LLM on an NVIDIA H200, including model inference and SID decoding with metadata probes and beam selection.

\begin{figure}[!tb]
\centering
\includegraphics[width=\columnwidth, trim={4 0 4 4}, clip]{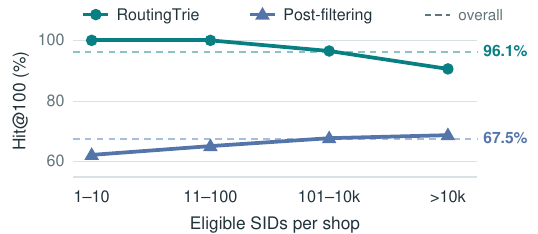}
\vspace{-20px}
\caption{\textbf{RoutingTrie improves retrieval across shop sizes.} Item level Hit@100 on 23,167 historical purchase sessions, grouped by the number of eligible SIDs for the target shop. Dashed lines show overall hit rates across all sessions.}
\label{fig:quality-by-group}
\vspace{-4px}
\end{figure}

\paragraph{Results.}\label{sec:real-results}
RoutingTrie improves item level Hit@100 by applying shop constraints before beam selection. Across all replay sessions, Hit@100 rises from 67.5\% with post filtering to 96.1\% with RoutingTrie, a gain of 28.6 percentage points. Figure~\ref{fig:quality-by-group} shows improvements in every shop size group. The largest gain occurs for shops with 1 to 10 eligible SIDs, where Hit@100 increases from 62.3\% to 100.0\%. The remaining groups gain 21.8 to 34.8pp. With identical model scores, RoutingTrie removes prefixes that cannot reach the requested shop before they consume beam capacity, helping retain purchased targets when eligible SIDs are scarce. In the deployment measurement, RoutingTrie with Sparse List changes median GR request latency from 27.1 to 28.2 ms, with an increase of only 1.1 ms. %

\section{Limitations}
\label{sec:discussion}

RoutingTrie indexes categorical metadata, so numerical predicates such as price ranges require discretization before routing. Since a bin may include ineligible values, resolved items must still be checked against the original predicate. General Boolean requests require additional care when composing edge summaries, and their effect on item level retrieval quality remains unevaluated. 

Our reported results are based on the most common production design for GR systems, using residual codebooks and a catalog trie. Systems with independent codebooks or other SID constructions can follow the same routing principle when using a trie, but their performance may differ. Also with a finite beam in real GR system, prefixes that pass the edge masks but contain no eligible items may displace prefixes containing eligible items, so RoutingTrie cannot guarantee lossless retrieval at the item level.

\section{Conclusion}
\label{sec:conclusion}

In this paper, we introduced RoutingTrie, a simple yet effective solution to apply metadata constraints during SID generation in GR before beam selection. It builds compact field indexes over one shared trie and uses their probe results to mask candidate edges, helping preserve beam capacity for eligible items. Its Dense Bitset and Sparse List indexes share one trie while balancing memory use and probe speed across metadata fields. We also provide an analytical storage cost rule to guide index choice across fields with different cardinalities and value distributions. Experiments on ShopTrie demonstrate compact storage and efficient CPU and GPU probing. In a production replay, RoutingTrie improves Hit@100 by 28.6 percentage points, with only 1.1 ms latency increase. These findings support incorporating metadata constraints into the generation process as a practical way to improve eligible retrieval with modest resource costs.

\bibliography{references}
\bibliographystyle{icml2025}

\end{document}